# FOCAL MECHANISM UNCERTAINTY QUANTIFICATION IN GROUND MOTION SIMULATIONS OF LETEIL EARTHQUAKE

Valeria Soto[1], Fernando Lopez-Caballero[2]

[1]Postdoctoral researcher, Université Paris-Saclay, CentraleSupélec, ENS Paris-Saclay, CNRS, LMPS - Laboratoire de Mécanique Paris-Saclay, 91190, Gif-sur-Yvette, France (valeria-paz.soto-moncada@centralesupelec.fr)

[2]Professor, Université Paris-Saclay, CentraleSupélec, ENS Paris-Saclay, CNRS, LMPS - Laboratoire de Mécanique Paris-Saclay, 91190, Gif-sur-Yvette, France (fernando.lopez-caballero@centralesupelec.fr)

**ABSTRACT**

Ensuring the seismic safety of nuclear power plants (NPPs) is essential, especially for facilities that rely on base isolation to reduce earthquake impacts. For understanding the seismic response, accurate models are key to predict the ground motions, which are generally sensitive to various factors, including earthquake source parameters like the focal mechanism, i.e., strike, dip, and rake angles. This study examines how uncertainties in these parameters affect ground motion predictions. The analysis is based on the SMATCH benchmark, which provides a standardized approach for evaluating the seismic response of the Cruas-Meysse NPP in France during the Mw 4.9 Le-Teil earthquake of 2019. A set of 27 3D high-fidelity numerical simulations was performed using a spectral-element method, each incorporating different focal mechanism variations. These simulations provide an effective approach for investigating the factors behind the exceptional ground motion observed during this event. To quantify uncertainty, the simulated ground motions were compared to recorded data using two well-established goodness-of-fit criteria: one assessing time-frequency domain characteristics and another focusing on the characterization of the ground motion signals by intensity measures. Results highlight the significant influence of focal mechanism variability on ground motion predictions, especially on the rake angle, which showed the strongest correlation with wave and intensity measures.

**INTRODUCTION**

Ensuring the seismic safety of nuclear power plants (NPPs) is a fundamental priority, which can be assessed by seismic hazard. A main challenge is the consideration of uncertainty in ground motion prediction, which plays a crucial role in estimating how an earthquake's energy propagates and affects structures. Seismic hazard analysis is inherently uncertain, with both epistemic (knowledge-based) and aleatory (random) uncertainties influencing risk assessment outcomes (Tarbali et al., 2018). Among the different factors, earthquake rupture forecast uncertainties play a key role (Bradley et al., 2012), including earthquake source parameters by their focal mechanism variability (strike, dip, and rake angles). This factor is often simplified in hazard models, despite its potential to alter seismic wave propagation characteristics.

The Mw 4.9 Le Teil earthquake in 2019 provides a case study highlighting the consequences of focal mechanism uncertainty. Despite its moderate magnitude, the event produced exceptionally strong ground shaking, exceeding peak accelerations typically expected for similar events (Causse et al., 2021). The discrepancy between observed and predicted ground motions raises questions about the adequacy of current seismic hazard models and their ability to capture focal mechanism effects on ground motion variability.

Recent advances in physics based simulations (PBS) have greatly improved the ability to model complex seismic phenomena with high fidelity, including variations in magnitude-distance relationships, source focal mechanisms, and fault rupture processes (Castro-Cruz et al., 2021). PBS has successfully replicated historical earthquakes, demonstrating its ability to match observed ground motions while also predicting possible future scenarios under different seismic conditions. According to Causse et al. (2021), the Le Teil event produced intense ground motion especially in the vertical direction, a feature that cannot be accurately reproduced with more simplified 1D or 2D models. Given these advantages, this study employs a spectral element method (SEM3D), a well-established PBS approach, to model ground motion variability for the Le Teil earthquake scenario.

The results are part of the efforts given within the SMATCH benchmark, a computational framework developed by IRSN, EDF, and OECD-NEA in France for evaluating seismic hazard assessments of the Cruas-Meysse NPP, located in south-eastern France, which was subjected to Le Teil earthquake. This research simulates ground motion propagation from the earthquake source to a recording station near the Cruas-Meysse NPP. A total of 27 ground motion simulations were conducted, systematically varying strike, dip, and rake angles to evaluate their impact on predicted ground motion characteristics. The goals are (i) to assess whether these variations explain discrepancies between simulated and recorded ground motions and (ii) to determine how they should be incorporated into seismic hazard modelling for critical infrastructure. For quantifying the variations, the study employs two goodness-of-fit (GOF) criteria: (i) Time-frequency domain GOF from Kristeková et al. (2009) and in terms of Intensity Measures (IMs) GOF from Anderson (2004).

**PHYSICS BASED SIMULATIONS (PBS) USING SEM3D**

**Model and mesh description**
To analyse the impact of focal mechanism uncertainty on ground motion predictions, 3D PBS were conducted using the SEM3D (Touhami et al., 2022). Simulations were performed within a high-resolution numerical domain measuring 86.3 km × 72.8 km × 29.2 km, designed to accurately capture wave propagation effects, as observed in Figure 1(a).

The model is composed by 2,598,452 elements, each containing five Gauss-Lobatto-Legendre (GLL) points. With a minimum element size of $\Delta L = 75$ m, the maximum attained frequency is of about $f_{max} = 12$ Hz. To prevent artificial reflections at the domain boundaries, absorbing boundaries are implemented using perfectly matched layers (PMLs), ensuring that outgoing seismic waves did not interfere with the modelled ground motions.

All simulations were carried out on the high-performance computing (HPC) facilities of the *Mésocentre de Calcul de l'Université Paris-Saclay*. Each simulation represented a 12-second earthquake event, a duration chosen to fully capture the key wave phenomena without extending the simulation cost. Each simulation required a real-time duration of approximately 7,543 seconds and was executed using 400 CPUs.

**Source modelling and variability**
To represent the Mw 4.9 Le Teil earthquake source in a simplified yet physically meaningful way, this study adopts a point-source approximation defined by its focal mechanism. In this framework, the rupture is characterized by three angular parameters: strike (θ), dip (δ), and rake (λ), which describe the orientation of the fault plane and the direction of slip during rupture (Aki & Richards, 2002). The seismic moment of $M_0 = 2.81 \times 10^{16}$ N·m and initial focal mechanism parameters were defined as θ =45°, δ=55° and λ=90°. Figure 1(b) illustrates the assumed fault geometry used in the simulations.

Variations in these parameters affect the rupture orientation and ground motion pattern. To quantify focal mechanism uncertainty, a set of 27 simulations was performed, systematically varying strike, dip, and rake within a ±5° to ±10° range. Moreover, the earthquake rupture process was simulated using a moment function based on Liu et al. (2006) or a rise time of 1 s, following the results for the PBS of the same earthquake in Smerzini et al. (2023).

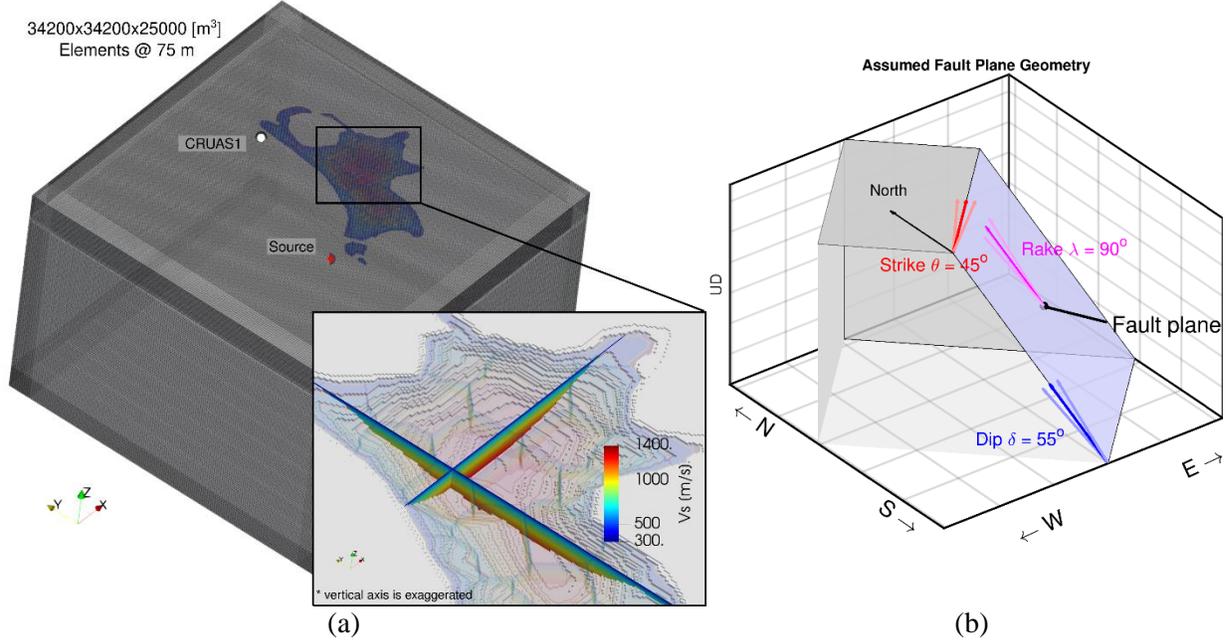

Figure 1. (a) Mesh description of the simulated ground motion model subjected to Le-Teil earthquake, including the regional and site conditions close to the Cruas-Meysse NPP. (b) Assumed fault plane geometry and focal mechanism. The strike (θ = 45°), dip (δ = 55°), and rake (λ = 90°) angles are represented in red, blue, and magenta arrows, respectively.

**Material properties**

This study incorporates the explicit modelling of basin geometry in the PBS framework. The Cruas-Meysse region contains complex subsurface geological structures that could significantly affect seismic wave propagation. To account for site effects, the simulation incorporates a 3D velocity model provided from Smerzini et al. (2023) for the basin sediments, and from Causse et al. (2021) for the regional crustal bedrock, which can be observed in Figure 1(a): for the deepest (crustal) layers, Table 1 summarizes the values used. Additionally, as observed in Figure 1(a), the S- and P-wave propagation velocities ($V_s$ and $V_p$, respectively) inside the sedimentary basin vary with depth $z$ according to the following relationships:

$$V_s(z) = 300 + 53.7\, z^{0.5} \quad (2)$$
$$V_p(z) = 550 + 78.3\, z^{0.5} \quad (3)$$

Table 2: Crustal model material properties (from Causse et al. 2021).

| Depth (m) | $\rho$ (kg/m3) | $V_p$ (m/s) | $V_s$ (m/s) | $Q_p$ | $Q_s$ |
|---|---|---|---|---|---|
| 0 | 2500 | 3366 | 2047 | 400 | 180 |
| 628 | 2600 | 5995 | 3645 | 400 | 180 |
| 1197 | 2300 | 1967 | 1200 | 400 | 180 |
| 1416 | 2500 | 3831 | 2291 | 400 | 180 |
| 2026 | 2500 | 3908 | 2314 | 400 | 180 |
| 2194 | 2600 | 5819 | 3457 | 400 | 180 |
| 5956 | 2600 | 5951 | 3616 | 400 | 180 |

**Validation with real data using GOF criteria**

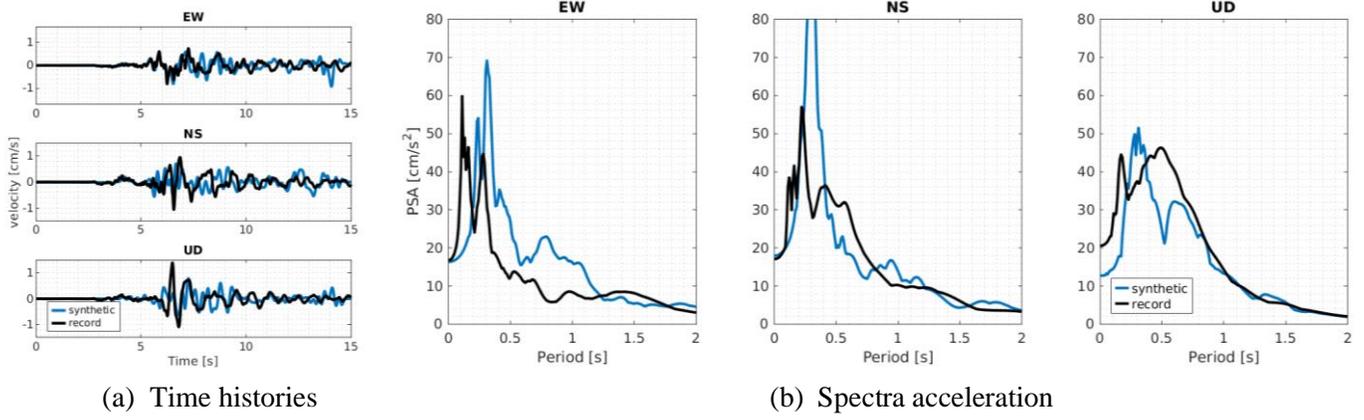

(a) Time histories    (b) Spectra acceleration

Figure 2. Comparison of synthetic (blue) and recorded (black) ground motion at station CRUAS1: (a) time histories and (b) acceleration response spectra.

To evaluate the accuracy of the physics based simulations (PBS), the synthetic ground motions were compared against recorded seismograms from station CRUAS1, which is located on the northern side of the fault, approximately at a distance of 15 km, as observed in Figure 1(a). This validation employed two well-established Goodness-of-Fit (GOF) criteria that assess waveform similarity (or disagreement) from complementary perspectives: time-frequency domain characteristics and intensity measures. In both cases, the GOF values fall into four quality levels: poor (1–4), fair (4–6), good (6–8), and excellent (8–10).

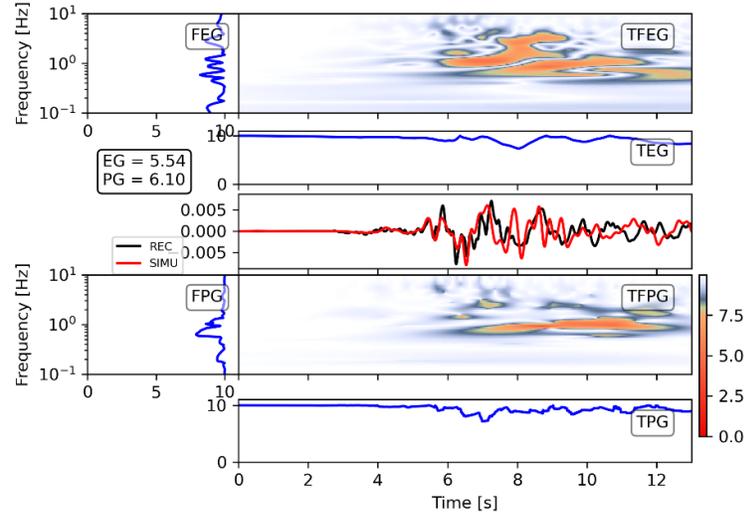

Figure 3. Summary of the GOF criteria in the time-frequency (TF) domain (Kristeková et al., 2009) applied to the recorded and synthetic data for station CRUAS1, in the EW component. The figure shows the TF misfit representation of envelope (TFEG) and phase (TFPG), along with the misfit in time (TEG, TPG) and frequency (FEG, FPG).

The first set is the method proposed by Kristeková et al. (2009) which analyses the agreement in waveform shape through amplitude (or envelope, EG) and phase (PG) comparisons, both in time and frequency domains. These values can be further decomposed into Time (TEG, TPG) and Frequency (FEG, FPG). Figure 3 exemplify the GOF misfit applied to the recorded and synthetic data for station CRUAS1, in the EW component. The obtained values: EG=5.54 and PG=6.10 show that the simulation attains a global fair-to-good agreement.

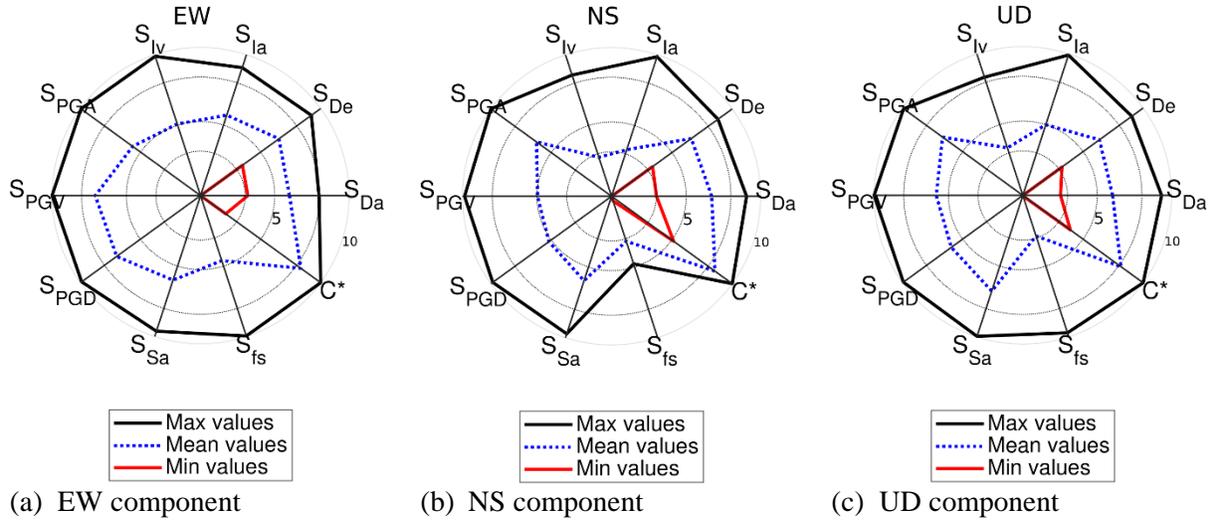

(a) EW component  (b) NS component  (c) UD component

Figure 4. GOF criteria using ground motion characterization by 10 intensity measures (IMs) (Anderson, 2004) applied to the recorded and synthetic data for station CRUAS1, in the EW component. The figure shows the maximum, mean and minimum values from 7 frequency bands between 0.05 and 10 Hz.

The second set follows the method of Anderson (2004) focusing on key intensity measures (IMs), which assess the similarity of two signals by using the characterization of the ground motion as a quantitative score. This characterization is performed using 10 intensity measures (IM) as scores: the peak ground acceleration (PGA), peak ground velocity (PGV), peak ground displacement (PGD), Arias intensity (Ia), Arias duration (Da), Energy duration (De), the integral of velocity squared (Iv), response spectra (Sa), Fourier spectra (fs) and cross correlation (C*). The computation is performed for 7 frequency bands from 0.05 to 10 Hz. Figure 4 show the IM-based GOF scores for each motion components. According to this GOF criteria, the similarity between the simulation and the recorded data is in fair-to-good agreement, the same as with the GOF in the time-frequency domain. The EW component has a global better agreement, with scores S>5 in every IM, while NS has the lowest agreement. These differences can be related to several factors, such as the use of a point source as a simplification, which is not able to account for complex factors like directivity in that direction.

## RESULTS AND DISCUSSION

To evaluate the impact of focal mechanism variability on ground motion predictions, an extensive set comprising a total of 27 high-resolution 3D simulations was performed using systematically varied combinations of strike ($\theta$), dip ($\delta$), and rake ($\lambda$) angles. Each simulation was evaluated against recorded data using two complementary Goodness-of-Fit (GOF) approaches. Specifically, we used the EG and PG scores from Kristeková's time-frequency framework and the mean values from Anderson's intensity measure-based criteria as presented above.

### Influence of focal mechanisms variability on Ground motion predictions

Figure 5 presents a concise summary of the results from the 27 simulations, grouped by fault parameter (strike, dip, rake) and ground motion component (EW, NS, UD), in order to compare how each source parameter influences simulation performance. Depending on the angle variation, some effects can be observed:

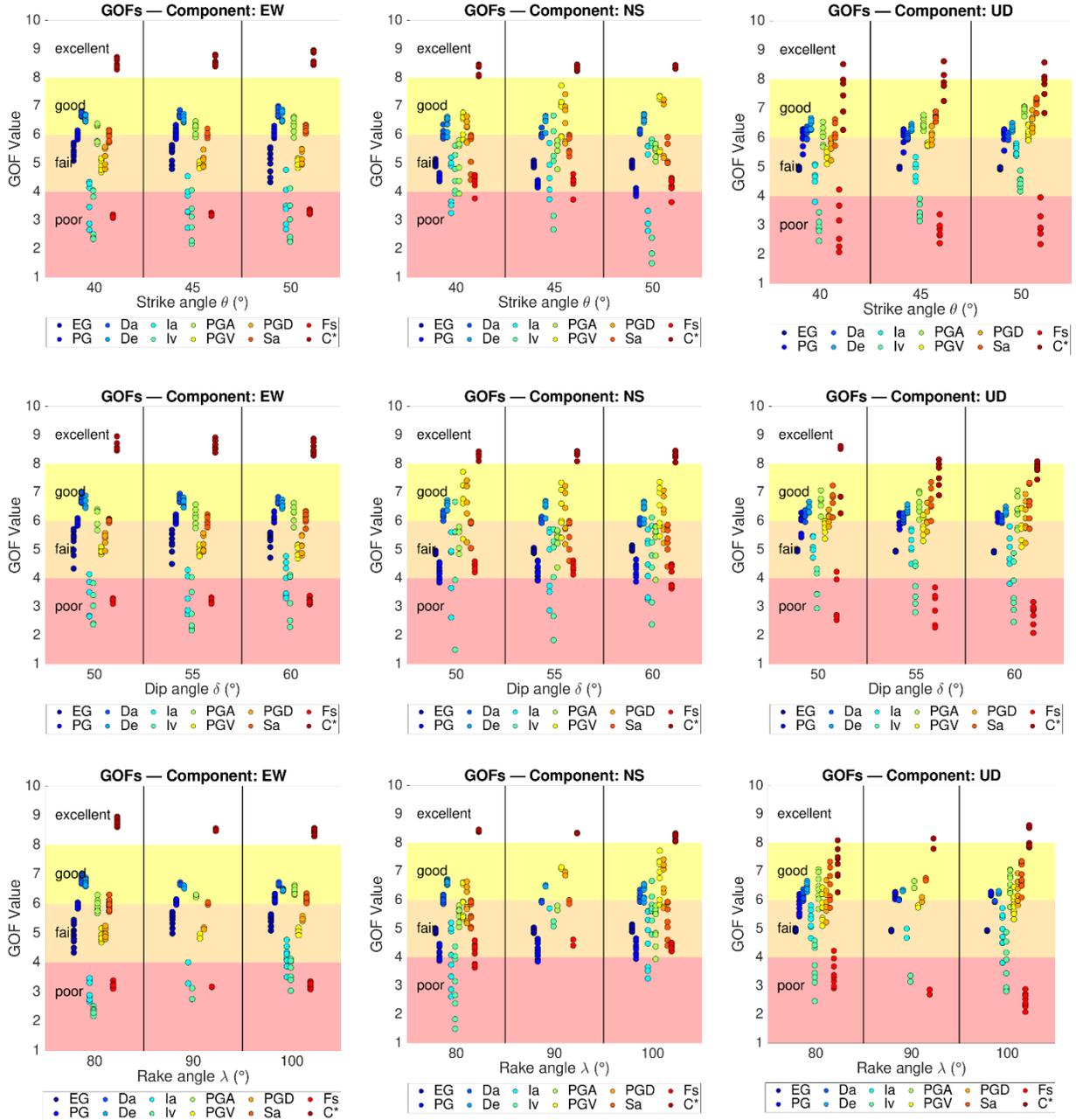

(a) EW component  (b) NS component  (c) UD component

Figure 5. Impact of focal mechanism variability on ground motion simulation performance: GOF scores grouped by fault parameter: strike (top), dip (middle) and rake (bottom). The GOF scores are grouped to represent the quality of the fit between simulated and recorded data, from red (poor fit) to yellow (fair to good fit) to white (excellent fit).

1. *Strike Angle Effects ($40°$, $45°$, and $50°$)*. In Figure 5(a), across the variation range, GOF scores remained stable for the UD and EW components. EG and PG values, in blue colours, mostly stayed in the fair to good range, which aligned well with measures like PGV, Sa, and C*, especially in the EW direction. This indicates that strike angle has limited influence on EW and UD motions. However, the NS component showed more scattered results with poor performance in several cases. This trend in the NS direction indicate that strike influences the ground motion across this axis.

2. ***Dip Angle Effects (50°, 55°, and 60°)***. In Figure 5(b), the UD component showed the clearest improvement in the agreement with increasing the dip angles, particularly with EG, PG, PGD, PGA, and Fs. The EW component remained stable, while the NS component again showed more scatter, with some IMs falling below the fair threshold. This suggests that dip angle affects waveform agreement primarily in the vertical direction, likely due to its control over rupture inclination and energy radiation toward the surface. While its overall influence is stronger than strike, dip alone does not explain the discrepancies in the NS component, pointing to additional contributing factors.
3. ***Rake Angle Effects (80°, 90°, and 100°)***. Among the three parameters, rake angle had the strongest influence on the GOF results, as observed in Figure 5(c). In the UD component, most EG and PG values increased at 80° and 100°, with PG generally performing better than EG. This improvement extended to Anderson's scores, where IMs like PGD, PGV, and especially C* reached the excellent range. The EW component followed a similar pattern, showing its best match at 80°, while scores at 90° and 100° were slightly lower but still within the good range. The NS component, though more scattered, also responded well to a rake of 100°, where scores improved across both GOF sets. The strong dependence of GOF scores on rake angle highlights its key role in shaping ground motion patterns. Rake controls the direction of slip during rupture; therefore, small variations have a direct impact on wave radiation and energy distribution.

In summary, while the two GOF criteria focus on different metrics, the results obtained from them are generally consistent. This suggest that when changes in source parameters lead to better waveform agreement, they also improve intensity measure agreement. Across all components, the EW direction obtained the best fit, indicating that it is less sensitive to source-modelling uncertainties. In contrast, the NS direction showed persistent disagreement, which may reflect the simulation's limited ability to reproduce ground motions in that direction. This could be attributed to several factors, such as geometric simplifications, the lack of complexity in the source representation due to the point-source approximation, especially in a direction where directivity effects may occur, or local site effects.

**Correlation Between Fault Geometry and Ground Motion Fit**
To better understand how focal mechanism parameters influence the GOF scores, correlation coefficients were calculated between the fault angles (strike, dip, rake) and each GOF metric. While correlation coefficients were computed to explore potential relationships between fault parameters and GOF metrics, it is important to note that each parameter (strike, dip and rake) was sampled with only three discrete values. Therefore, the results should be interpreted as qualitative trends. In this context, correlation values indicate the strength and direction of the relationship between a given parameter and a simulation score: positive values suggest that increasing the parameter leads to better agreement, while negative values imply that increasing the parameter worsens the fit. The three heatmaps in Figure 6 summarize these correlations for each motion component (UD, NS, EW). Only correlation values that are statistically significant ($p \leq 0.05$) are displayed, while non-significant results are omitted for clarity.

As observed in Figure 6(a), most values showing strong positive relationships in the EW direction are associate to the rake angle. Metrics such as Iv, PGD, and PGA exceeded 0.84, while De and C* also followed this trend but with a negative correlation. These high values confirm that even small variations in rake can enhance the agreement between simulations and observations, especially in terms of amplitude and energy. In contrast, the dip and strike angles showed fewer strong correlations. One exception was the link between strike and Fs (0.84), which suggests a connection between strike orientation and spectral content in this component.

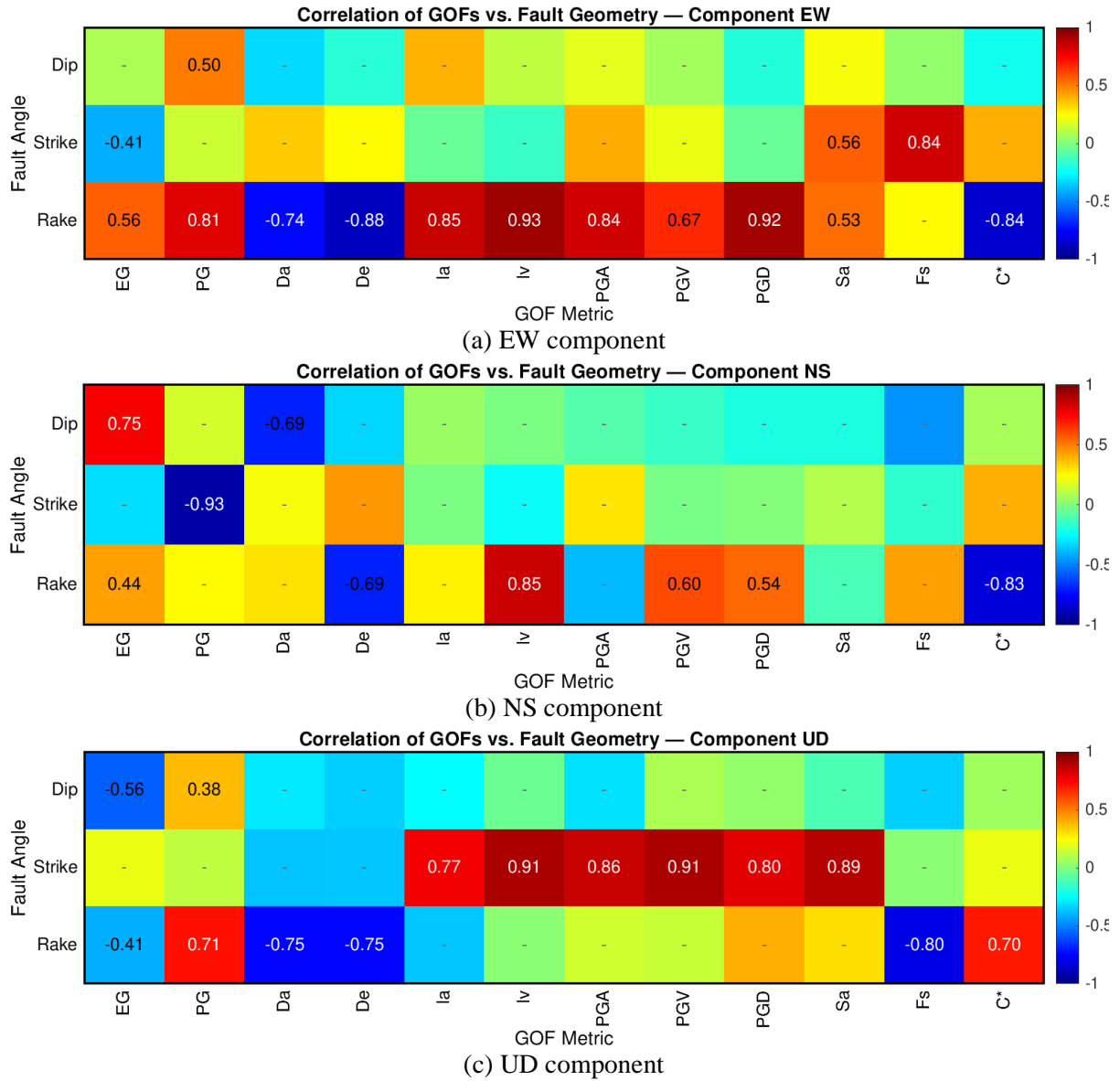

Figure 6. Correlation heatmaps between focal mechanism parameters (strike, dip, rake) and GOF scores (Kristeková and Anderson criteria) across components: (a) EW, (b) NS, and (c) UD. Only statistically significant values are shown.

Moreover, in the NS direction (Figure 6(b)), the patterns became less clear and more scattered. The strike angle had a particularly strong negative correlation with PG (–0.93), which shows that phase agreement worsens when the strike angle increases. This trend matches earlier observations from the GOF plots, where NS was more sensitive in the time-frequency domain, associated to its waveform. The dip angle, however, showed that EG increased with dip (0.70), while other Da decreased (around –0.70), which shows that while the overall energy distribution may align better with larger dips, duration metrics do not follow the same trend. Additionally, as well with the other horizontal direction, in the NS the rake angle has the greatest number of significant correlations, while they are moderate to strong positive correlations, especially for Iv and PGD (0.85 and 0.60, respectively). On the other hand, Fs and C* presented strong negative links. Therefore, these values reinforce that the NS component reacts differently depending on the type of metric and that it is especially sensitive to phase and duration mismatches.

Finally, for the vertical direction (Figure6(c)), the strongest correlations appeared in Anderson's IM-based scores. The strike angle had the most visible impact, showing strong positive relationships with nearly all IMs. Metrics such as Iv, PGV, Sa, and PGD reached correlation values above 0.85, indicating that when the strike angle increased, the fit of these measures improved. This trend confirms what was already observed in the GOF plots. Moreover, the rake angle also affected some IMs, especially Fs and C*, where it showed opposite effects. A strong negative correlation appeared for Fs (–0.80), while C* increased with rake (0.70). These values suggest that rake modifies the waveform shape in the frequency domain, even when intensity measures stay stable. The dip angle, in contrast, showed fewer strong correlations. The only clear trend was a negative link with EG (–0.62), suggesting that envelope misfit increases as the dip angle grows. Therefore, in the vertical component, strike appears to control the fit of intensity measures, while rake and dip have more localized effects on spectral and waveform features.

The correlation plots show that small changes in fault geometry, particularly the rake angle, can lead to noticeable shifts in the agreement between simulated motions and the observed data. For the UD and EW components, the rake angle had the strongest correlations with most intensity measures, especially PGV, PGD, Iv, and Sa. This means that the direction of slip on the fault plane plays a key role in shaping the amplitude and energy of the ground motion. In contrast, strike and dip angles had a more limited influence. Strike showed a strong link with some vertical motion metrics, but dip mostly affected envelope and duration-related measures. These results support the idea that focal mechanism uncertainty can significantly change simulation outcomes. Another observation comes from the NS component, which showed weaker and more scattered correlations across all fault parameters. This lower agreement could come from the modelling approach, since in this study the source was modelled as a point, which is a common simplification for moderate-magnitude events like Le Teil. However, a point source cannot capture complex rupture behaviours such as directivity, slip heterogeneity, or finite-fault effects, which are known to influence the directional distribution of energy (Spudich et al., 2019).

**CONCLUSION**

This study examined how variations in focal mechanism parameters affect ground motion predictions at a station located near the Cruas-Meysse NPP subjected to the 2019 Mw 4.9 Le Teil earthquake, using a set of 27 numerical simulations. The focal mechanism, defined by the strike, dip and rake, was systematically varied to evaluate its influence on simulated ground motions.

The results show that the rake angle has the strongest influence on both waveform shape and intensity measures, followed by strike and dip. This influence was most clearly observed in the UD and EW components, where high correlations were found between rake and key metrics such as PGD, PGV, and spectral amplitudes. The vertical component, which showed notably high peak accelerations during the Le Teil event, responded to changes in dip and strike, due to its role in rupture inclination and vertical energy radiation. However, these variations do not seem sufficient to fully explain the observed high PGA ground motion in this direction. Furthermore, the NS component showed a low agreement between record and simulations, regardless the focal mechanism.

These last two observations may be explained by limitations of the point-source approximation, which does not include source complexities such as the rupture directivity or spatial variability in slip. These effects are known to strongly influence wave propagation, particularly in near-source conditions or in regions with complex geological structures (Spudich et al., 2019).

While these complexities were not explicitly modelled in this study, the results still provide valuable insight into the sensitivity of ground motion predictions to focal mechanism variability, enabling a controlled assessment of source uncertainty. As observed, relying on a single focal mechanism in hazard analysis may overlook realistic variations that significantly affect ground motion outcomes. Including this variability, especially for critical sites such as nuclear power plants, could improve the robustness of seismic assessments and help better capture the range of expected motions (Bradley et al., 2012; Tarbali et al., 2018).

**ACKNOWLEDGMENTS**

This work was performed using computational resources from the "Mésocentre" computing center of Université Paris-Saclay, CentraleSupélec and École Normale Supérieure Paris-Saclay supported by CNRS and Région Île-de-France (https://mesocentre.universite-paris-saclay.fr/)

**REFERENCES**

Aki, K., & Richards, P. G. (2002). *Quantitative seismology*.

Anderson, J. G. (2004). Quantitative measure of the goodness-of-fit of synthetic seismograms. *13th World Conference on Earthquake Engineering Conference Proceedings, Vancouver, Canada, Paper*. http://www.iitk.ac.in/nicee/wcee/article/13%7B%5C_%7D243.pdf

Bradley, B. A., Stirling, M. W., McVerry, G. H., & Gerstenberger, M. (2012). Consideration and propagation of epistemic uncertainties in New Zealand probabilistic seismic-hazard analysis. *Bulletin of the Seismological Society of America*, *102*(4), 1554–1568. https://doi.org/10.1785/0120110257

Castro-Cruz, D., Gatti, F., & Lopez-Caballero, F. (2021). High-fidelity broadband prediction of regional seismic response: a hybrid coupling of physics-based synthetic simulation and empirical Green functions. *Natural Hazards*, *108*(2), 1997–2031. https://doi.org/10.1007/s11069-021-04766-x

Causse, M., Cornou, C., Maufroy, E., Grasso, J. R., Baillet, L., & El Haber, E. (2021). Exceptional ground motion during the shallow Mw 4.9 2019 Le Teil earthquake, France. *Communications Earth and Environment*, *2*(1). https://doi.org/10.1038/s43247-020-00089-0

Kristeková, M., Kristek, J., & Moczo, P. (2009). Time-frequency misfit and goodness-of-fit criteria for quantitative comparison of time signals. *Geophysical Journal International*, *178*(2), 813–825.

Liu, P., Archuleta, R. J., & Hartzell, S. H. (2006). Prediction of broadband ground-motion time histories: Hybrid low/high-frequency method with correlated random source parameters. *Bulletin of the Seismological Society of America*, *96*(6), 2118–2130.

Smerzini, C., Vanini, M., Paolucci, R., Renault, P., & Traversa, P. (2023). Regional physics-based simulation of ground motion within the Rhône Valley, France, during the MW 4.9 2019 Le Teil earthquake. *Bulletin of Earthquake Engineering*, *21*(4), 1747–1774. https://doi.org/10.1007/s10518-022-01591-w

Spudich, P., Cirella, A., Scognamiglio, L., & Tinti, E. (2019). Variability in synthetic earthquake ground motions caused by source variability and errors in wave propagation models. *Geophysical Journal International*, *219*(1), 346–372. https://doi.org/10.1093/gji/ggz275

Tarbali, K., Bradley, B. A., & Baker, J. W. (2018). Consideration and propagation of ground motion selection epistemic uncertainties to seismic performance metrics. *Earthquake Spectra*, *34*(2), 587–610. https://doi.org/10.1193/061317EQS114M

Touhami, S., Gatti, F., Lopez-Caballero, F., Cottereau, R., de Abreu Corrêa, L., Aubry, L., & Clouteau, D. (2022). SEM3D: A 3D High-Fidelity Numerical Earthquake Simulator for Broadband (0–10 Hz) Seismic Response Prediction at a Regional Scale. *Geosciences (Switzerland)*, *12*(3), 0–27. https://doi.org/10.3390/geosciences12030112